\documentclass[12pt]{revtex4} \usepackage[dvips]{graphicx,color} \def\bea{\begin{eqnarray}}
\usepackage{amsmath}
\def\eea{\end{eqnarray}}
\def\be{\begin{equation}}
\def\ee{\end{equation}}

\def\k{\kappa}
\def\k{\kappa}

\def\m{\mu}
\def\n{\nu}

\begin{document}
\title{Signs and Cosmology\footnote{Essay Written for Gravity Research Foundation's 2012 Awards for Essays on Gravitation}}

\vspace{1cm}

\author{\sc Arundhati Dasgupta and Adamantia Zampeli}
\affiliation{
  Department of Physics and Astronomy, University of Lethbridge, Lethbridge, Canada T1K 3M4}
\email{arundhati.dasgupta@uleth.ca; a.zampeli@uleth.ca}
\begin{abstract} 

Using a new effective action predicted from quantum gravity where the sign of the Euclidean Einstein action is reversed
we discuss WKB solutions in quantum cosmology and predict that this sign change might explain the origin of phantom scalars.
\end{abstract}
\maketitle
\section{Introduction}
It is well known that Euclidean gravity has a problem: The Euclidean Einstein action is unbounded from below. This implies that the action can be arbitrarily negative
for certain metric configurations. Hawking identified this problem \cite{haw} of
the action and attributed it to the fact that the gravitational potential is attractive and thus negative. However, it is believed that
gravitational energy measured asymptotically is positive and this led to the positive action conjecture for Euclidean gravity
proved in some situations \cite{yau}.
An explanation why attractive gravity eventually has positive energy and hence positive action is again due to Hawking, who observed that matter eventually collapse
to form black holes and once the black hole is formed, its energy in the form of mass is positive. 
However the positive energy conjecture is yet to be proved for generic metrics and the problem becomes very crucial in the evaluation of the Euclidean path-integral. It was obvious that
in the Euclidean path-integral defined thus
\be
Z=\int {\cal D} g_{\m \n} e^{ -I_{\rm Eucl}}
\ee
would be large positive weights from the Boltzmann factor if the gravitational action $I_{\rm Eucl}$ were negative and path-integral would be divergent even before 
the integral was computed.
Several authors, including one of us, motivated from \cite{schleichhartle} tried to show that the measure in the path-integral is
non-trivial and thus what one should consider is the `effective action' \cite{lolladg,mazur,adg1} in the computation of the path-integral. This will 
include the contribution from the
Faddeev-Popov determinant, and thus
\be
Z=\int {\cal D} g^{\rm phys}_{\m \n} e^{-I_{\rm meas}-I_{\rm Eucl}}
\ee
would be the physical path-integral, which might be well defined. Recently, due to observations from \cite{lolladg,adg1,loll} this seems to be correct.
In the weak gravity regime the effective action is simply a sign reversed form of the original Euclidean action. This seems to be particularly important, as it is 
confirmed from independent Lattice gravity (Causal Dynamical Triangulation) evaluations
of the path-integral for the action for the scale factor or cube root of the three volume of the spatial slices which is evolved in time \cite{loll}. 
The question therefore which naturally arises is whether this change in sign is relevant or irrelevant if one Wick rotates back to Lorentzian space. This is what we will investigate in this essay in the context of quantum cosmology.
\section{The Reversed Sign}
In \cite{adg1} it was shown that the contribution of the measure in the Euclidean path-integral in a certain approximation
simply rescales the contribution from the classical action. This rescaling prefactor is a function of the undetermined
constant $C$ which determines the signature of the DeWitt supermetric \cite{adg1}. This result was a confirmation of a similar
observation from Causal Dynamical Triangulation computation \cite{loll}. To be precise,
\be
I_{\rm Eucl}= -\frac{1}{16\pi G}\left[1+ \frac{2(1+2C)}{\pi}\right]\int \sqrt{g_{\rm Eucl}} (R-2\Lambda)\ d^4 x_E \label{scale}\ee
For C=-2 which is Einstein-Hilbert gravity, the prefactor is positive and we can therefore write the prefactor
as the square of a real number, say $b^2$.
If we Wick rotate this rescaled Euclidean action, the Lorentzian action obtained from the above (according to the prescription of \cite{haw}),
the new redefined Lorentzian action $I'_{\rm Lor}$ is thus 
\be
I'_{\rm Lor}=-i I'_{\rm Eucl}= - i b^2\int \sqrt{g_{\rm Lor}}(R-2\Lambda)\ d^4 x_E = -b^2 \int \sqrt{-g_{\rm Lor}}(R-2\Lambda) \ d^4 x\ee
Thus the sign of the Lorentzian action is also reversed. 

\be
I'_{\rm Lor}= -\frac{b^2}{\kappa} I_{\rm Lor}
\ee
where $\kappa=1/16\pi G$ and $b^2= -\frac{1}{16\pi G}\left[1+ \frac{2(1+2(-2))}{\pi}\right]$. 
To see how this might affect a typical system let us take the minisuperspace metric for cosmology:
\be
ds^2=- N^2 dt^2 + a^2(t)\ d\Omega
\ee
where $a(t)$ is the scale factor of the spatial three slices and $d\Omega$ the metric on the three slice and $N$ the lapse.
The three metric has curvature proportional to k=-1,0,1 representing negative, no and positive curvature.
The classical Lagrangian can be derived for this metric:
\be
I_{\rm Lor}= p_a \dot a -N \left(\frac{-p_a^2}{24 a}-6 k a\right)
\ee
($p_a$ is the momentum conjugate to $a(t)$).
Given this, it is obvious that if we use the rescaled action to find the momenta $p'_a$,
\be
p'_a= \frac{\partial I'_{\rm Lor}}{\partial \dot a}=-\frac{b^2}{\k}p_a
\ee
and thus the modified Hamiltonian $H'$ is 
\be
H'= p_a'\dot a - I'_{\rm Lor}= -\frac{b^2}{\kappa}H
\ee  
The dynamics of this system is usually studied as coupled to a scalar field and thus the total Hamiltonian is
\be
-\frac{b^2}{\kappa} H + H_{\rm matter}
\label{modified}
\ee
due to obvious reasons, we assume that the matter Hamiltonian does not scale.
Thus the total Hamiltonian seems to differ non-trivially from the original one.
We then examine how this modified Hamiltonian might affect the physics in two different cases (i) the WKB wavefunction and tunneling \cite{vilenkin}
in quantum cosmology and (ii) detect a particular way of defining phantom scalar fields.

\section{Cosmology and the WKB}
The total Hamiltonian for an FRW model with a quantum scalar field is \cite{isham}
\begin{equation}
H = -\frac{p_a^2}{24a} -6ka +\frac{p_\phi^2}{2a^3} + a^2 V(\phi)
\end{equation}

For the most simple case where $k=1$ and $V(\phi)=0$, i.e. the quantum field is massless and for a choice of time $t=a(t)$, the Hamiltonian constraint takes the form

\begin{equation}
-\frac{p_a^2}{24t}-6t +\frac{p_\phi^2}{2t^3} =0 \Rightarrow p_a^2 = 24t (-6t +\frac{p_\phi^2}{2t^3}) \label{squaredmomentum}
\end{equation}
and the true (squared) Hamiltonian is 
\begin{equation}
h_a^2 \equiv p_a^2 = -144t^2 +\frac{12p_\phi^2}{t^2} 
\end{equation}
The squared true Hamiltonian is a function of time and has 4 roots, two real and two imaginary,
\begin{align}
t_{1,2} &= \pm \frac{\sqrt{p_\phi}}{12^{1/4}}\\
t_{3,4} &= \pm i \frac{\sqrt{p_\phi}}{12^{1/4}}
\end{align}
Analysis of the sign shows that in the interval between the two real roots, $(t_1,t_2)$ we have $h_a^2 > 0$, and thus it is a classically allowed region, while in the intervals from $(-\infty, t_1)$ and $(t_2,+\infty)$ it is negative and $h_a$ becomes imaginary. Thus, we can have quantum tunneling.

We can see this using the WKB approximation method for the Schrodinger equation
\begin{equation}
p_a \psi (t) = i \hbar \frac{\partial}{\partial t} \psi (t) 
\end{equation}
The solutions to this equation in the WKB approximation are of the form
For $p^2_a >0$
\begin{equation}
\psi (t) = \frac{c}{\sqrt{p_a}} e^{\pm \frac{i}{\hbar} \int_{t_1}^{t_2} dt p_a(t)}
\end{equation}
+ is the expanding, - is the contracting case.\\
For $p^2_a <0$
\begin{equation}
\psi (t) = \frac{d}{\sqrt{|p_a|}} e^{\pm \frac{1}{\hbar} \int dt |p_a(t)|}
\end{equation}
where $c,d$ constants to be defined by the imposition of boundary conditions.\\
In the case the cosmological part of the Hamiltonian is multiplied by a factor $\mu=-b^2/{\kappa}$ as in (\ref{modified})
\begin{equation}
H= (-\frac{p_a^2}{24a} -6ka) \mu +\frac{p_\phi^2}{2a^3} + a^2 V(\phi)
\end{equation}
which we let to be either positive or negative as per the sign of C in (\ref{scale}). Again for $k=1$, $V(\phi)=0$ and choice of time $a(t)=t$, the true squared Hamiltonian is 
\begin{equation}
h_a^2 \equiv p_a^2 = -144a^2 + \frac{12 p_\phi}{2 \mu a^2}
\end{equation}
The roots of this Hamiltonian have the form
\begin{align}
t_{1,2} &= \pm \frac{\sqrt{p_\phi}}{{(12\mu)}^{1/4}} \label{root1}\\
t_{3,4} &= \pm i \frac{\sqrt{p_\phi}}{{(12\mu)}^{1/4}} \label{root2}
\end{align}
which again are two real and two imaginary. In the case $\mu>0$, the extended Hamiltonian is positive in the interval between the real roots and negative otherwise, and for the imaginary time, in the interval between the two roots it is negative and positive otherwise, while when $\mu<0$ as in the case for Einstein Gravity which has C=-2  the real and imaginary roots switch. 
The form of the solutions to the Schrodinger equation in the WKB approximation will again have the same form as previously, but now the wavefunctions  will depend on the parameter $\mu$. The general solution for all the regions, classically allowed and not allowed is
\begin{equation}
\psi (t) = \frac{c}{\sqrt{|p_a(t, \mu)|}} e^{\pm \frac{i}{\hbar} \int_{t_1}^{t_2} dt p_a(t, \mu)}
\end{equation}
Depending on whether $\mu$ is positive or negative we recover the solution in each interval.
As the real and Imaginary roots are degenerate this does not change the nature of the WKB wavefunctions nor the tunneling. Thus even though the `Euclidean effective action' is no-longer unbounded, the physics remains unchanged. If the degeneracy in the roots is broken by e.g. invoking a non-zero $V(\phi)$, 
there will be a distinct change in the turning points due to change in the sign of the action. This is described
in the Appendix. Though this is a quantitative change in the location of the turning points or the tunneling process, 
qualitatively the system is not really affected.\\
\section{Phantom of the Cosmology}
Even though apparently the standard FRW cosmology is qualitatively unchanged, there is one aspect where this simple change in sign
can bring new physics, which is in the coupling of gravity to phantom fields. These scalar fields are usually taken to be those
which have a negative kinetic term in the Lagrangian and thus give rise to negative pressure systems (\cite{carrol} and references therein) and thus dark energy.  
We begin by observing that the typical gravity and matter coupled system has the Lagrangian:
\be
\int \sqrt{-g} \left [\kappa R - \frac12\partial_{\mu}\phi g^{\m \n}\partial_{\nu}\phi - V(\phi)\right]\  d^4 x
\ee
(remember $\kappa=1/16\pi G$).
If we then introduce the `scaling' due to the quantum corrections in the gravity sector, this modifies to:
\be
\int \sqrt{-g} \left[-b^2 R - \frac12\partial_{\mu}\phi g^{\m \n}\partial_{\nu}\phi - V(\phi)\right]\ d^4 x
\ee

 We extract the minus sign as an overall signature of the system
\be
-\int \sqrt{-g}  \left[ b^2 R + \frac12\partial_{\mu}\phi g^{\m \n} \partial_{\nu} \phi + V(\phi)\right] \ d^4 x
\ee
If we then analytically continue in the scalar field $\phi'=i \phi$ and assume that $V(\phi)= g\phi^4/4!$ The Lagrangian transforms to
\be
-\int \sqrt{-g} \left[ b^2 R -\frac12\partial_{\m}\phi'g^{\m\n}\partial_{\n}\phi' + V(\phi')\right]\ d^4 x
\ee

Needless to say that this is slightly different from the usual phantom fields where the Kinetic term changes sign, the potential remains as it is. What we have achieved here 
is a `relative'
change of sign between the kinetic term of the potential and the other terms in the coupled Lagrangian. Infect if we compute the density and the pressure of this 
scalar matter
\bea
\rho & =& -\frac{{\dot\phi}^{'2}}{2} + \frac{V(\phi')}{2}\\
p&=&-\frac12\left[{\dot\phi}^{'2} + V(\phi')\right]
\eea
And thus $w=p/\rho <-1$ even for this re-defined cosmology \cite{carrol}!
This precisely identifies phantom fields, however this model has to be investigated further. For example, we can choose as an internal time the field $\phi$ and then quantize the system. Then, starting from the observation that, in that case, the analytic continuation in the field becomes a Wick rotation, we can study the behaviour of the quantized system in the realms of quantum cosmology.

\section{Conclusion}
Thus for Euclidean gravity the reversal of the sign of action due to quantum correction makes a big difference path-integral computation 
which becomes a convergent integral. However
it does not affect the `classical' and semiclassical physics of the system per se. Quantum corrections which produce higher curvature terms 
also found in \cite{adg1} in the strong gravity limit we expect that will affect the dynamics of the system. 
We examined the case for WKB approximations for the tunneling wavefunction in quantum cosmology and the results remained qualitatively 
unchanged from those predicted using the original Lagrangian.
 Surprisingly, however, we are able to interpret
the change in sign in the gravitational sector of the theory to provide a plausible origin of `phantom negative pressure'.
This method of creating negative pressure is however confined to scalar potentials which are mapped to themselves
under such an analytic continuation. We have to investigate this further for any deeper implications and hope to report on this in the near future.

\appendix
\section{Non-Degenerate turning points}
General form of the solutions for 
\begin{equation}
p_a^2 = 24 (-6t^2 + \frac{p_\phi^2}{2 t \mu} + \frac{t^4 V(\phi)}{\mu})
\end{equation}

\begin{align}
t_{1,2} &= \pm \frac{\sqrt{2}}{2V(\phi) M^{1/3}}[V(\phi) M^{1/3} K]^{1/2}\\
t_{2,3} &= \pm \frac{1}{2 V(\phi) M^{1/3}} [V(\phi) M^{1/3} (-K + 12 \mu M^{1/3} + i \sqrt{3} (M^{2/3} -16 \mu^2))]^{1/2}\\
t_{5,6} & =\pm \frac{1}{2 V(\phi) M^{1/3}} [V(\phi) M^{1/3} (-K + 12 \mu M^{1/3} - i \sqrt{3} (M^{2/3} -16 \mu^2))]^{1/2}
\end{align}
where
\begin{align}
M &= -2 p_\phi^2 V^2(\phi) \pm 64 \mu^3 + 2p_\phi V(\phi)\sqrt{p_\phi^2 V^2 (\phi) \mp 64 \mu^3}\\
K&= M^{2/3} + 16 \mu^2 \pm 4 \mu M^{1/3}
\end{align}
and the upper sign corresponds to $\mu>0$, while the lower sign to $\mu<0$.
Thus a rescaling of the Lagrangian due to quantum corrections does change the turning points, though qualitatively the system remains the same.


\end{document}